\documentclass[aps,prl,superscriptaddress,twocolumn]{revtex4}
\usepackage{graphics}
\usepackage{graphicx}
\usepackage{color}

\begin{document}

\title{Signatures of non-Abelian statistics in non-linear coulomb blockaded transport}%
\author{Roni Ilan}
\affiliation{Department of Condensed Matter Physics,
Weizmann Institute of Science, Rehovot 76100, Israel}
\author{Bernd Rosenow}
\affiliation{Institut f\"ur Theoretische Physik, Universit\"at
Leipzig, D-04009 Leipzig, Germany}
\author{Ady Stern}
\affiliation{Department of Condensed Matter Physics, Weizmann
Institute of Science, Rehovot 76100, Israel}

\date{\today}

\begin{abstract}
Signatures of the non-Abelian statistics of quasi-particles in the
$\nu=5/2$ quantum Hall state are predicted to be present in the
current-voltage characteristics of tunneling through one or two
quantum Hall puddles of Landau filling $\nu_a$ embedded in a bulk
of filling $\nu_b$ with $(\nu_a,\nu_b)=(2,5/2)$ and
$(\nu_a,\nu_b)=(5/2,2)$.
\end{abstract}


\maketitle

Anyons obeying Non-Abelian statistics\cite{MooreRead} were
predicted to exist in condensed matter systems almost twenty years
ago. Although many efforts have been made to suggest and perform
experiments to test that hypothesis, there is still no clear-cut
proof for the existence of such
particles\cite{RevModPhysTQC,SternPedRev}. This work proposes an
experimental setup along with a set of measurements designed to
unravel signatures of such statistics in the putative non-Abelian
quantum Hall state at $\nu=5/2$. In particular, we show that the
current-voltage characteristics of a Coulomb blockaded system in
the non-linear regime may detect the neutral Majorana edge mode
which is an essential property of the proposed non-Abelian state.
Moreover, it allows us to study the spectrum of this mode, and
explore the effect of the presence of quasiparticles on that
spectrum.

Coulomb blockade effects in the regime of linear response were
suggested before as a tool to probe non-Abelian phases
\cite{SternHalperin,ilan-2007,ilan-2008}, via tunneling of
electrons into large dots. Coulomb blockade peaks are observed
when the ground state of a dot with $N$ electron is degenerate
with that of a dot with $N+1$ electrons. Thus, their study probes
a ground state property as a function of parameters such as the
size of the dot and the magnetic field.  In contrast,
current-voltage characteristics in the non-linear regime of the
Coulomb blockade hold information of the many particle excitation
spectrum of the system. A peak in the differential conductance
$dI/dV$ will appear whenever a proper resonance condition between
the source-drain voltage and the excitations of the dot is met.

\begin{figure}[b]
\includegraphics[width=1\linewidth]{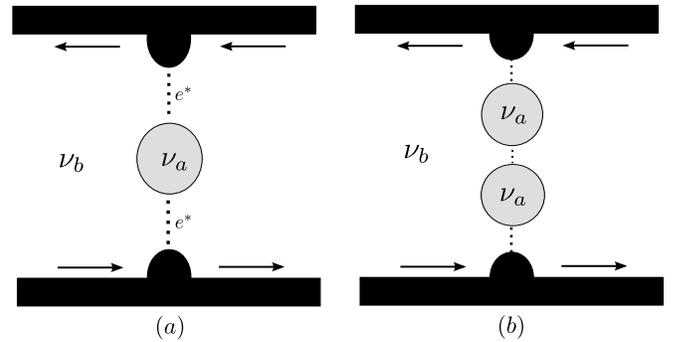}
\caption{Antidots in a Hall bar geometry. The bulk filling
fraction $\nu_b$ determines the tunneling excitation, and the
antidot's filling fraction is $\nu_a$. (a) A single antidot. (b)
Double antidot geometry, discussed towards the end of the paperh.
}\label{fig:setup}
\end{figure}

In this work we analyze $dI/dV$ in a Coulomb blockade measurement
as a function of source-drain voltage and magnetic field. We show
that the "diamond structure" that characterizes the Coulomb
blockade out of the linear response regime could serve to identify
the nature of the $\nu=5/2$ state. We consider a Hall bar or a
Corbino disk in which a quantum anti-dot, a puddle of filling
fraction $\nu_a$, is  embedded in a bulk of filling factor $\nu_b$
(see Fig.~\ref{fig:setup}). The puddle is surrounded  by a gapless
region. Since this region is compact, its spectrum is quantized.
Transport from one edge of the sample to the other is facilitated
by tunneling of charge carriers (whose precise nature is discussed
below) through the anti-dot, and
 is characterized by resonances corresponding to the
internal states of the gapless region around the antidot. The
edges are connected to reservoirs and their spectrum is
continuous. We assume that the anti-dot is weakly and
symmetrically coupled to the nearby edges, and that tunneling
through the anti-dot can be treated in the sequential tunneling
approximation.

In the setup we propose, the type of charge-carrier allowed to
tunnel across the bulk is determined by the bulk filling fraction
$\nu_b$. For an integer $\nu_b$ only electrons are allowed to
tunnel, while a fractional $\nu_b$ allows for quasi-particles of
fractional charge to tunnel as well. Electron tunneling
spectroscopy in the quantum Hall regime has been achieved before
for integer filling fraction in Ref.~\cite{GoldmanAntiDot}. The
experiment we consider probes the spectrum and set of states
available to the tunneling excitation in the edge separating the
$\nu=5/2$ phase from the integer $\nu=2$ phase. Below, we first
consider the case $\nu_b=5/2, \nu_a=2$, where our setup should in
principle be easier to stabilize with a top-gate due to the
robustness of the the integer $\nu=2$ plateau\footnote{In fact,
our results hold also when $\nu_a=1$ or $0$}. We then continue
with the combination $\nu_a=5/2, \nu_b=2$.

The various descriptions suggested for the $5/2$ phase describe
the energy of the edge separating the anti-dot from the bulk as
consisting of a sum of several contributions, one arising from the
charge mode and the others arising from neutral modes, should
those exist. The ground state energy is obtained when that sum is
the lowest. The charging energy has the standard form
\begin{equation}\label{eq:charging}
E_c=\frac{\pi v_c}{\alpha L}\left(N-N_\phi-N_g\right)^2
\end{equation}
where $v_c$ is the charge velocity, $L$ is the perimeter of the
dot, and $N$ counts the number of charge-carriers on the edge.
Here, $N_\phi$ and $N_g$ are shifts to the effective number of
charge-carriers due to extra flux or the presence of back-gate
voltage. If $N$ counts the number of electrons, $\alpha=1/\nu=2$,
while if $N$ counts the number of quarter-charged quasiparticles,
$\alpha=8$.

The existence and spectrum of the neutral modes vary between
different theories of the $\nu=5/2$ state. For concreteness, we
first focus on the Moore-Read theory\cite{MooreRead} and explore
the features it will show in experiment. Then we extend the
discussion to include some other known models for the edge, and
show that the predictions they lead to are different from those of
the Moore-Read state.


For the case of $\nu_b=5/2$ in a Moore-Read state and $\nu_a=2$,
quasi-particles of charge $e/4$ will hop onto the edge of the
anti-dot, thereby charging it as well as creating neutral
excitations on it. The neutral edge mode in this case was
extensively studied (See for example Ref.\cite{ReadGreen}) and
shown to be described by a single chiral Majorana fermion theory.

The spectrum of the Majorana edge fermions varies with the number
of quasi-particles on the edge. The Hamiltonian is of the form
\begin{equation}
H=\frac{2\pi v_n}{L}\left [\sum_{n\geq0}\epsilon_n c_n^\dagger c_n
+\Delta\right ] \label{neutralspectrum}
\end{equation}
where $c_n^\dagger, c_n$ are fermionic creation and annihilation
operators, while $v_n$ is the velocity of the excitations. A
recent numerical work~\cite{PffafianNumerics} estimates the
neutral velocity $v_n$ to be between $0.75\times 10^6$ and
$1.1\times 10^6$ cm/s for GaAs in the thermodynamic limit, and
$v_c$ to be between $6$ and $8$ times larger. For an even number
of quasi-particles, the single fermion states are of energy
$\epsilon_n=n+1/2$ and $\Delta=0$, while for an odd number of
quasi-particles, the available single fermion states are of
energies $\epsilon_n=n$ and $\Delta=\frac{1}{16}$. This difference
in the single fermion spectrum leads to a pronounced difference in
the many fermion spectrum of the Hamiltonian
(\ref{neutralspectrum}). For an odd number of quasi-particles, the
spectrum remains composed of non-negative integer multiples of
$2\pi v_n/L$. In contrast, for an even number of quasi-particles
the spectrum is composed of integer (when the number of fermions
is even) or half integers (when that number is odd) multiples of
$2\pi v_n/L$. Remarkably, in the latter case one integer is absent
from the spectrum. To get a total energy of $2\pi v_n/L$ from half
integer single fermion energies two fermions must occupy the state
of $\epsilon_n=1/2$, and that is forbidden by the Pauli principle.

These two different many fermion excitation spectra should
manifest themselves in the $I-V$ characteristics of the anti-dot
as a function of magnetic field (or the voltage on the gate
defining the antidot) and source-drain voltage $V$, the voltage
between the two edges. In the limit when $e^*V$ is well below
twice the charging energy of the anti-dot (with $e^*$ being the
charge of the tunneling charge-carrier), the flow of current takes
place through an alternation of the number of quasi-particles on
the anti-dot between two values, $N=n_{qp}$ and $N=n_{qp}+1$. A
quasi-particle that tunnels into or out of the anti-dot excites
its edge. If the anti-dot is weakly coupled to the two edges, the
current through it is small, and in between two consecutive
quasi-particles tunneling events the anti-dot relaxes to its
ground state. Thus, a tunneling process always starts with the
anti-dot in its ground state and ends with it being excited. In
view of the symmetry of the coupling between the anti-dot and the
two edges, the electrochemical potential of the anti-dot is
half-way between those of the edges.

For current to pass through the anti-dot, the minimal voltage
$V_{min}$ should supply enough energy to overcome the difference
between ground state energies $E_{gs}$ of $n_{qp}$ and $n_{qp}+1$
quasi-particles on the anti-dot,
\begin{equation}
e^*V_{min}/2=|E_{gs}(n_{qp}+1)-E_{gs}(n_{qp})|.
\label{diamondedge}
\end{equation}
For $V>V_{min}$ resonances in $dI/dV$ should occur when
\begin{eqnarray}\label{eq:condition1}
&&E(n_{qp}+1,N_g)-E_{gs}(n_{qp},N_g)=e^*V/2\\
&&E(n_{qp},N_g)-E_{gs}(n_{qp}+1,N_g)=e^*V/2.\label{eq:condition2}
\end{eqnarray}
where $e^*=e/4$ is the charge of the tunneling quasi-particle. The
first condition (\ref{eq:condition1}) is the one allowing
tunneling onto the dot. The voltage bias between the edge and the
anti-dot has to compensate for the energy difference between the
ground state of the dot with $n_{qp}$ quasiparticles, and an
excited state of the anti-dot with $n_{qp}+1$ quasiparticles.
Similarly, the second condition (\ref{eq:condition2}) allows
tunneling from the anti-dot onto the opposite edge.

Since our focus of interest is the neutral mode whose typical
energy scale is much smaller than that of the charged mode, we
will consider voltages small enough to keep the charged mode
always in its ground state. The tunneling quasi-particle may then
excite the neutral mode to any excited state to which its energy
is sufficient. The excitability of many fermion states of the
neutral mode may be understood by considering the description of
the Moore-Read state as a $p$-wave super-conductor of composite
fermions, and the neutral Majorana mode as the edge mode that
characterizes such a super-conductor\cite{ReadGreen}. The
tunneling quasi-particle is a vortex in this super-conductor.
When a $e/4$ quasi-particle tunnels through the bulk and onto the
anti-dot, it may break some of the pairs in the condensate.
Therefore, pairs or individual fermions may be carried with the
quasi-particle onto the anti-dot to occupy the single particle
energy modes of the Hamiltonian (\ref{neutralspectrum}).

\begin{figure}[t]
\includegraphics[width=0.48\linewidth]{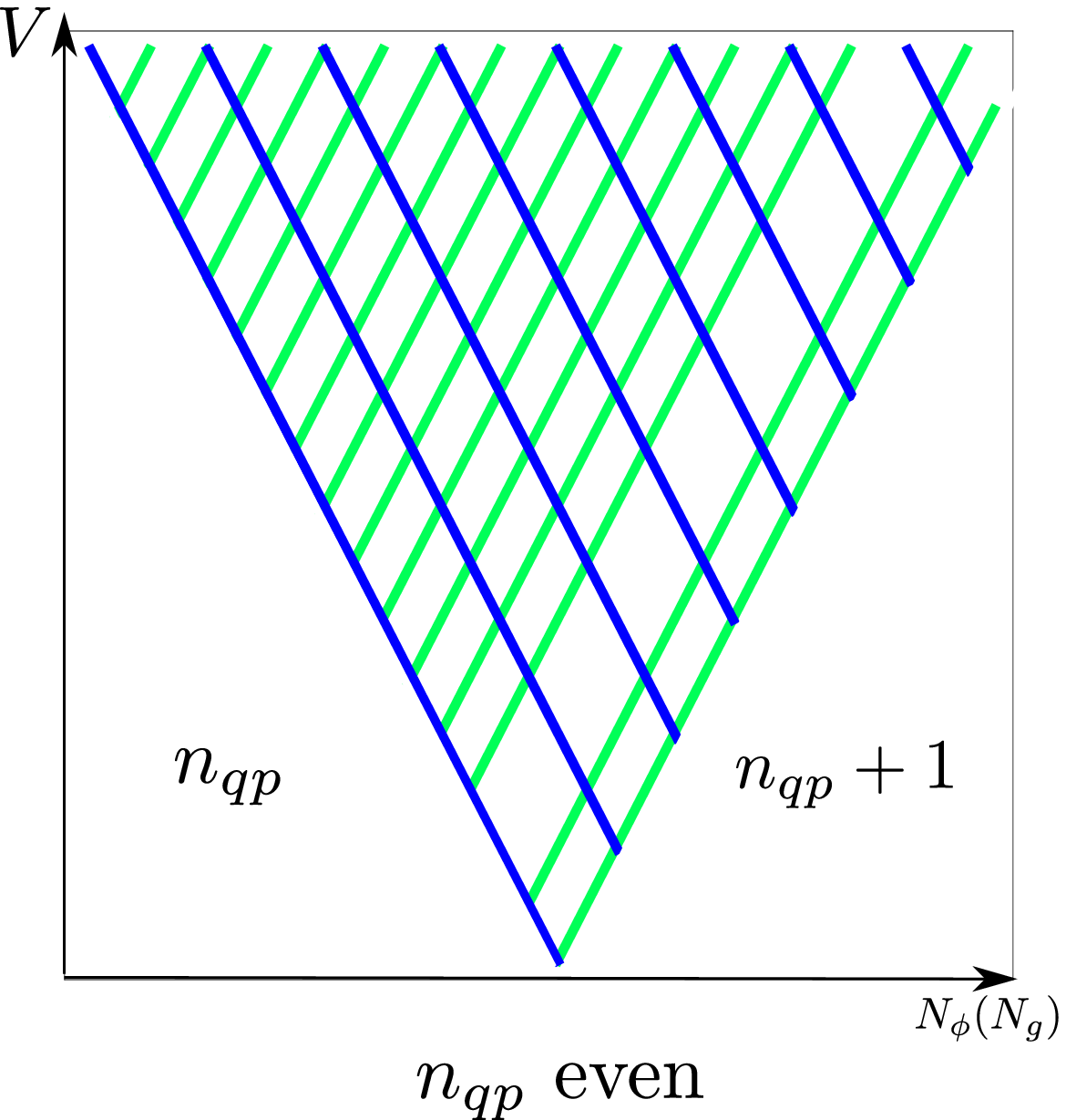}
\includegraphics[width=0.48\linewidth]{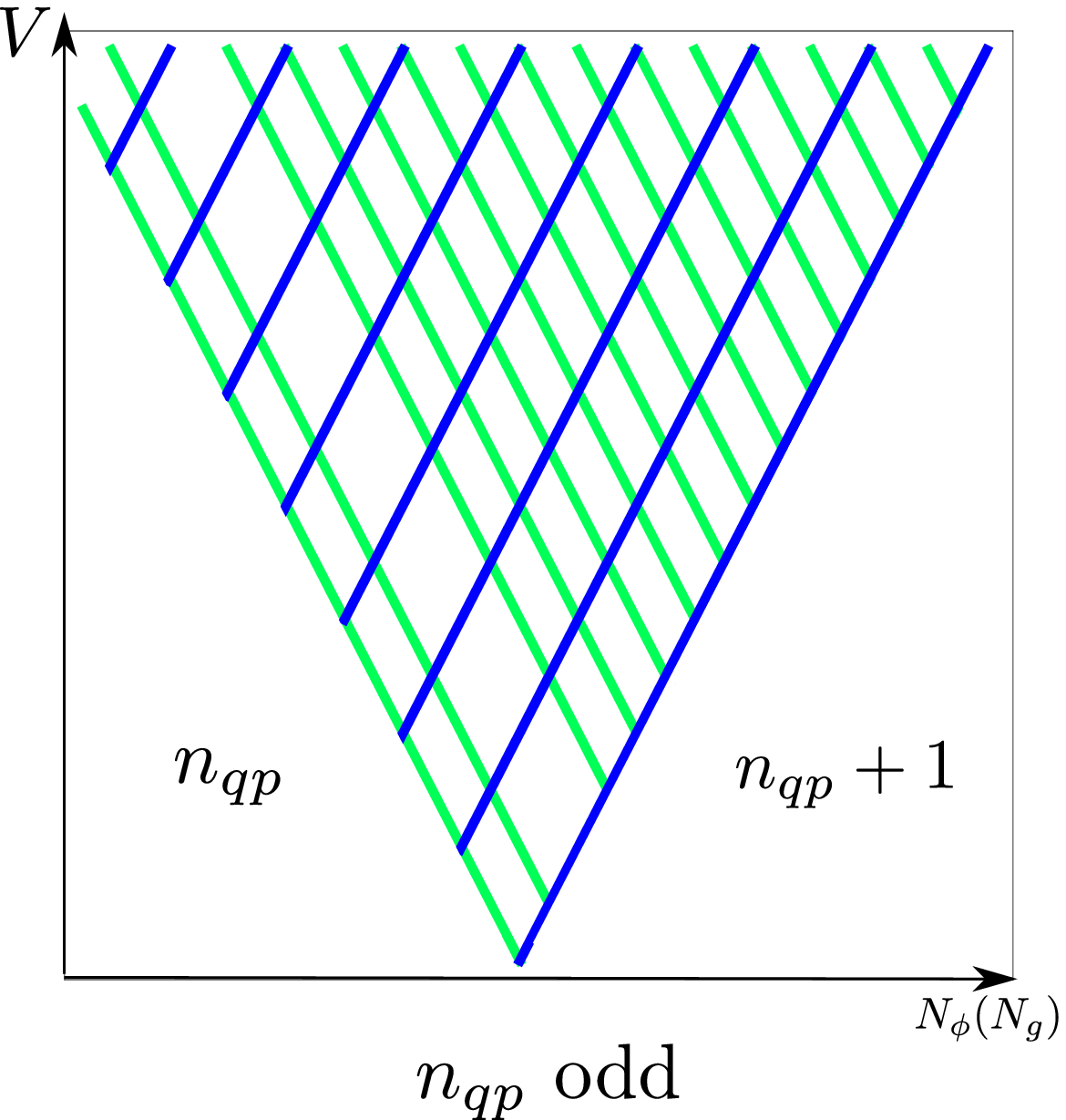}
\caption{Coulomb blockade diamond shapes in $dI/dV$ for a
transition between $n_{qp}$ and $n_{qp}+1$ quasi-particles on the
dot. White region represent a stable state of a dot with $n_{qp}$
or $n_{qp}+1$ quasiparticles (Coulomb blockaded regions). Green
lines are separated by half the distance separating blue lines.
}\label{fig:diamonds}
\end{figure}

With this picture of the tunneling process, we can now find the
solutions to Eqs. (\ref{eq:condition1}) and (\ref{eq:condition2}).
For $n_{qp}$ even, $E(n_{qp})=E_c(n_{qp})+2\pi v_nq/L $, and the
non-negative $q$ can obtain integer and half-integer values, with
the condition $q\ne 1$. For $n_{qp}$ odd,
$E(n_{qp})=E_c(n_{qp})+2\pi v_nq/L +2\pi v_n\Delta/L$, and $q$ can
obtain only integer values. Using equation (\ref{eq:charging}), we
find that resonances in $dI/dV$ as a function of  $V$ and $N_\phi$
appear as lines parallel to the edges of the Coulomb diamonds
defined by (\ref{diamondedge}). Plots of the location of resonance
for the two possibilities appear in figure~(\ref{fig:diamonds}).
As can be seen in the figure, the difference between the even-odd
and the odd-even transition from $n_{qp}$ to $n_{qp}+1$ is
manifested in the density of the lines of positive and negative
slopes (tunneling-out and tunneling-in resonances). When $n_{qp}$
is even, lines of positive slope are twice as dense as those of
negative slope. When $n_{qp}$ is odd, the converse is true - the
density of lines with a negative slope is twice as large as that
of lines with positive slope. This doubling of the number of
resonances reflects the difference in the many particle spectrum
of the Majorana fermion edge mode - the energy separation between
excited states is halved when changing $n_{qp}$ from odd to even.
Note that the second excitation line is missing on the denser set
of lines, reflecting the absence of an excitation of energy $2\pi
v_n/L$.

Now, let us consider the case where $\nu_a=5/2$ and $\nu_b=2$.
This case is rather similar to the strong backscattering limit of
a Fabry-Perot interferometer discussed in
Refs.~\cite{SternHalperin,ilan-2007,ilan-2008}. The charge on the
entire $\nu_a=5/2$ anti-dot (bulk+edge) is quantized in units of
the electron charge, but the charge on the edge may become
fractional, and quantized in units of $e/4$. This happens by the
creation of a quasi-particle/quasi-hole pair, with one member of
the pair located in the bulk of the anti-dot and the other on its
edge. The edge spectrum depends on the parity of the number of
these pairs, in a similar way to the spectrum we discussed above.
The tunneling object is an electron.  As the magnetic field is
varied the number of electrons that minimizes the energy of the
anti-dot varies, and so does also the number of edge
quasi-particles. Equations (\ref{eq:condition1}) and
(\ref{eq:condition2}) still describe the conditions for the onset
for tunneling, with $n_{qp}$ replaced by $N$, the number of
electrons on the edge, and $e^*$ replaced by $e$. Since the parity
of the number of $e/4$ quasiparticles on the edge of the anti-dot
is not affected by electron tunneling, the energy spectrum of the
neutral modes is the same for $N$ and $N+1$. It may be varied only
through changing the magnetic field or the gate voltage. Hence,
there should be no difference between the density of resonance
lines of positive and negative slope in $dI/dV$. The density of
these lines should be doubled/halved whenever a change in flux or
voltage modifies the number of quasiparticles localized inside the
anti-dot, and changes its parity.

We now turn to show how the experimental set-up that we discuss
may distinguish between different candidate states for $\nu=5/2$,
due to their different edge structure. We remind the reader that
the structure of lines for the Moore-Read state, seen in Fig.
(\ref{fig:diamonds}), results from a combination of three
ingredients; the existence of a neutral slow mode creates the
resonances, the dependence of its spectrum on the parity of
$n_{qp}$ leads to the two different densities of peaks, and having
just a single Majorana mode leads to the absence of a state of
energy $2\pi v_n/L$ for even $n_{qp}$.


Although it has been shown numerically that under certain
conditions the Pfaffian state faithfully describes the true ground
state for $\nu=5/2$, \cite{PffafianNumerics} so far only the
charge of the tunneling quasi-particle has been measured and
verified as $e/4$ \cite{Dolev,Marcus,Willett}. This is consistent
with several other candidates, such as the anti-Pfaffian
\cite{AntiPf1,AntiPf2,AntiPffNumerics}, the $331$ Halperin state
\cite{331}, the $K=8$ state and the $U(1)\times SU_2(2)$ state
\cite{su2wen,su2block}. All these candidate states are made of
condensed pairs of particles with the resulting quasi-particle
charge being $e/4$. We now examine the $I-V$ characteristics to be
expected for these states.

\textit{{The anti-Pfaffian:}} The random edge of the anti Pfaffian
supports one charge mode, and three counter propagating neutral
Majorana fermion modes\cite{overbosch-2008}. In the infra-red
limit, all three modes have the same velocity. There are three
different quasi-particle operators defined on that edge, and the
presence of a quasihole changes the boundary conditions for all
three at once. As a consequence of having several flavors of
majorana fermions, the analysis we carried out for the Moore-Read
state is changed only in one respect: the Pauli principle does not
prevent here the formation of an excitation of energy $2\pi v_n/L$
for an even $n_{qp}$. We note, however, that a finite quantum
anti-dot does not necessarily reach the infra-red limit. If it
does not, the velocities of the three Majorana modes will not
necessarily be identical, and the spectrum of resonances will be
more complicated.

\textit{{The 331 Halperin state:}} The edge of the $331$ state
carries a single chiral complex dirac fermion\cite{ReadGreen},
which is equivalent to two Majorana fermions. The Hamiltonian of
its neutral mode is then similar to Eq. (\ref{neutralspectrum}),
but with two flavors of fermions. The results are therefore
expected to be similar to those obtained for the Anti-Pfaffian
state. We note, however, that the $331$ state correspond to a
special symmetry point, of zero spin polarization. Any deviation
from this point will lead to a different set of resonances.

\textit{The K=8 state:} This state does not carry a neutral mode,
and hence does not give rise to the resonances we discuss here.

\textit{${SU(2)_2\times U(1)}$:} The edge structure of this state
is detailed in the appendix of Ref.~\cite{overbosch-2008}. The
edge accommodates a charge density mode, a spin density mode of
velocity $v_s$, and a Majorana mode. Due to the presence of the
spin mode, the separation between the resonance lines in the
differential conductance may become non-uniform. If $v_n\ll v_s$,
a similar structure to that predicted for the Moore-Read state may
emerge, otherwise a more complicated structure is expected.

We now turn to discuss resonant tunneling through a double
anti-dot geometry (see Fig. (\ref{fig:setup})). In this setup,
charge-carriers have to tunnel through both anti-dots to get
across the Hall bar. We first assume that the two anti-dots are at
$\nu=2$, and are Coulomb blockaded in such a way that current
flows by tunneling of quasi-particles in the sequence
$(m_{qp},n_{qp})\rightarrow (m_{qp}+1,n_{qp})\rightarrow
(m_{qp},n_{qp}+1)\rightarrow (m_{qp},n_{qp})$, with the two
numbers signifying the number of quasi-particles on the first and
second anti-dots, respectively. Furthermore, we assume that the
coupling between the anti-dots is weak enough such that the first
anti-dot relaxes to its ground state before the quasi-particle
tunnels on to the second one. Then, tunneling from one edge to the
other requires that the ground state of the first anti-dot is
aligned with an excited state of the second~\cite{DoubleDotRMP}.
This allows us to probe the doubling of the many particle density
of states of the second anti-dot as a function of the number of
quasiparticles its edge accomodates. This time, however, resonance
peaks should appear in the current $I$, rather than in the
differential conductance $\frac{dI}{dV}$. For voltages small
enough not to affect $n_{qp}$, the density of the peaks as a
function of gate voltage or source-drain voltage will reflect the
many-body spectrum of the neutral mode of the second anti-dot,
whose density of states depends on the parity of $n_{qp}$. The two
possible parities will be reflected in two different densities of
resonance peaks as a function of gate voltage or source-drain
voltage. Similarly, two different densities of peaks should be
observed also when the tunneling charge-carrier is an electron.

Before concluding, we comment on an essential ingredient of our
analysis, namely the requirement for relaxation of the neutral
edge mode(s) of the anti-dot(s) between tunneling events. A
mechanism for such relaxation is the coupling of the edge of the
anti-dot to quasiparticles localized in the bulk of the $\nu=5/2$
phase~\cite{ilan-2008}. This coupling increases with the density
of the bulk quasiparticles, and with their proximity to the edge
of the anti-dot. Consequently, the rate of relaxation of the
anti-dot due to this coupling may be controlled by the magnetic
field, which controls the density of bulk quasiparticles. For the
experiments we propose here, the width of the excited states of
the neutral mode of the anti-dot generated by their relaxation
rate to the bulk, should be smaller than their separation, yet
larger than the width corresponding to the rate at which
charge-carriers pass through the anti-dot. With the relaxation
rate being controllable by the magnetic field and the rate at
which charge-carriers pass being controllable by the coupling of
the anti-dot(s) to the point contact, the desired limit may be
obtained.  Remarkably, the upper limit on the relaxation rate is
far less stringent than that required for the realization of
earlier proposals~\cite{ilan-2008}.

To summarize, we explored a new geometry for the detection of
neutral edge modes in the quantum Hall effect through resonant
tunneling experiments. Such experiments at $\nu=5/2$, involving
quasi-particle tunneling or electron tunneling, can show unique
features for the non-Abelian Moore Read state.

The authors thank C. M. Marcus for useful discussions. AS and RI
were supported by the US-Israel Binational Science Foundation, the
Minerva foundation, and Microsoft's Station Q. BR acknowledges
financial support from the BMBF.


\end{document}